\documentclass[10pt,journal]{IEEEtran}
\ifCLASSINFOpdf

\else
\fi
\usepackage{graphicx,amsmath,amssymb,cite}
\usepackage{color}
\usepackage{amsfonts}
\usepackage{amsmath}
\usepackage{graphicx}
\usepackage{amssymb}
\usepackage{stfloats}
\usepackage{subfig}
\usepackage{mathrsfs}
\usepackage{algorithm}
\usepackage{algorithmicx}
\usepackage{algpseudocode}
\usepackage{amsmath}
\usepackage{booktabs}
\usepackage[numbers,sort&compress]{natbib}
\usepackage{comment}
\usepackage{arydshln}

\begin{document}

\title{FL-CsLoc-DH: A Federated Learning-Based Crowdsensing Localization under Dual Heterogeneity}

\author{Xin Cheng,~\IEEEmembership{Member,~IEEE}, Binxu Wang, Zixuan Ai, Yuqing Yang, Yu He, Feng Shu,~\IEEEmembership{Senior Member,~IEEE},
    and Guangjie Han,~\IEEEmembership{Fellow,~IEEE}

    \thanks{Xin Cheng, Binxu Wang, Zixuan Ai, Yuqing Yang, Yu He, and Guangjie Han are with the College of Information Science and Engineering, Hohai University, Changzhou 213200, China. (e-mail:xincstar23@163.com). (Corresponding author: Xin Cheng.)}
    \thanks{Feng Shu is with the School of Information and Communication Engineering, Hainan University, Haikou 570228, China.}}
\maketitle

\begin{abstract}
    Mobile Crowdsensing (MCS) provides a scalable way to collect indoor fingerprints and reduce professional calibration effort. To protect the location privacy of participating clients, Federated Learning (FL) is used so that raw fingerprints and measurement trajectories remain local. However, owing to the open and uncontrollable nature of MCS, participating clients collect fingerprints in different propagation environments and possess unequal labeling availability. These conditions jointly cause data and annotation heterogeneity, which degrades FL-based crowdsensing localization. In this work, we propose FL-CsLoc-DH, a federated semi-supervised learning framework for crowdsensing localization under dual heterogeneity. FL-CsLoc-DH coordinates Two-Stage Noisy Gradient Rejection (TS-NGR), Soft Fisher Confidence Alignment (S-FCA), and Trajectory-Smoothed Model Averaging (T-SMA) throughout localization model learning. TS-NGR suppresses unreliable client updates caused jointly by environment dependent data shifts and noisy pseudo labels. S-FCA limits parameter increments dominated by unlabeled data, thereby alleviating annotation heterogeneity during local training. T-SMA smooths the aggregated global localization model trajectory to reduce drift induced by data heterogeneity. Experiment results show that the three mechanisms provide complementary improvements in localization accuracy under the MCS environment while retaining raw client data locally.
\end{abstract}

\IEEEpeerreviewmaketitle

\section{Introduction}

Indoor positioning is a key enabler for context-aware mobile services, asset tracking, emergency response, and intelligent Internet of Things (IoT) applications \cite{zafari2019survey}. Compared with received signal strength, Wi-Fi Channel State Information (CSI) exposes fine-grained amplitude and phase responses over subcarriers and therefore supports richer spatial fingerprints \cite{yang2013rssi}. Representative systems have demonstrated that CSI can enable decimeter-scale localization \cite{kotaru2015spotfi} and can be effectively modeled by deep fingerprinting architectures \cite{wang2015deepfi,wang2016csi,wang2017biloc,wang2017cifi,chen2017confi,wang2018deep}. More recent work has reduced human calibration effort \cite{tong2020csi} and strengthened CSI representations through attention augmented residual learning \cite{zhang2023csi}. Nevertheless, constructing such fingerprint databases requires collecting and labeling fingerprints at many reference positions \cite{cheng2025ris}, which is time-consuming and laborious. Mobile Crowdsensing (MCS) alleviates this calibration burden by outsourcing fingerprint collection to decentralized ordinary users \cite{gufran2023fedhil}. However, centralizing massive crowdsourced raw CSI data and continuous user trajectories inevitably raises severe privacy leakage concerns, creating a critical deployment challenge \cite{sun2026toward}.

Federated Learning (FL) offers a natural foundation for privacy-preserving crowdsensing localization because clients can retain raw fingerprints locally and exchange only localization-model updates \cite{cheng2022edge}. Existing studies have investigated FL-based localization. FedLoc established a general federated framework for cooperative localization and location-data processing \cite{yin2020fedloc}, while pseudo-label-driven crowdsourcing \cite{li2020pseudo} and federated Wi-Fi fingerprinting \cite{nagia2022federated} showed how decentralized users can contribute to indoor positioning models. Subsequent studies improved aggregation through model uncertainty \cite{park2022federated}, addressed heterogeneous fingerprint databases \cite{cheng2022federated}, introduced graph-based personalization \cite{wu2022multi} and online semi-supervised personalization \cite{wu2022prediction}. And some works extended FL-based localization to multibuilding environments \cite{gao2022federated}, multifloor environments \cite{guo2023fedpos}, mobile crowdsensing deployments \cite{gufran2023fedhil}, spatially heterogeneous CSI environments \cite{harada2026adafed}, and non-independent and non-identically distributed (non-IID) fingerprint environments with crowdsourced unlabeled data \cite{hussain2025semisupervised}. Collectively, these works validate the feasibility of federated indoor localization, but they do not address an MCS scenario in which clients simultaneously collect CSI fingerprints in different propagation environments and have different location label annotation ratios.

A realistic CSI crowdsensing deployment therefore exhibits dual heterogeneity. First, CSI fingerprints are collected by clients in spatially separated regions. Because each region has a different wireless propagation environment, fingerprint and coordinate data in different clients follow non-identical distributions \cite{cheng2022federated,gao2022federated,guo2023fedpos,gufran2023fedhil}. This condition is referred to as data heterogeneity. Second, location label annotation ratios in clients differ from each other because coordinates are available only at some reference positions. A fast-moving client may collect many unlabeled fingerprints, whereas a slow-moving client may perform more labeled measurements. This condition is referred to as annotation heterogeneity. These two factors constitute the dual data heterogeneity (i.e., data heterogeneity and annotation heterogeneity) addressed in this work. Furthermore, variations in device mobility and capabilities, such as computation and battery capacity, constitute practical crowdsensing deployment conditions. These conditions affect CSI stability, local computation, and participation reliability, thereby intensifying the effects of the dual heterogeneity.

Federated Semi-Supervised Learning (FSSL) exploits unlabeled client data by generating pseudo labels, but incorrect pseudo labels can reinforce their own errors during local training. Existing FSSL methods address this problem through model consensus \cite{liang2022rscfed}, class balanced pseudo labeling \cite{li2023class}, selective knowledge assimilation \cite{cho2023local}, multi-player generation \cite{che2021fedtrinet}, private learning \cite{fan2022private}, unlabeled client aggregation \cite{elbatel2024learning}, confidence discrepancy handling \cite{liu2025mind}, and few label robustness \cite{lee20242}. Related centralized Semi-Supervised Learning (SSL) studies also utilize curriculum pseudo labeling \cite{zhang2021flexmatch}, self adaptive thresholds \cite{wang2022freematch}, and soft confidence weighting \cite{chen2023softmatch}. These methods mainly use class probabilities and decision thresholds, which have no direct equivalent in continuous coordinate regression. Small coordinate errors also have different spatial meanings at different positions. Federated Fisher Focus Filtering (FedF3) studies annotation heterogeneity using adaptive loss filtering and Fisher-based parameter selection \cite{peng2025navigating}. However, CSI localization is a continuous regression problem, and hard rejection can discard spatially rare but useful client information. Moreover, a dynamically growing crowdsensing dataset changes client sample counts and pseudo-label quality during training.

Data heterogeneity causes local fingerprint and coordinate distributions to differ, so local localization model gradients point toward different local optima. Directly averaging these updates can make the global localization model drift or oscillate. Dynamic regularization \cite{acar2021federated}, model contrastive alignment \cite{li2021model}, sharpness aware optimization \cite{qu2022generalized}, and discrepancy aware collaboration \cite{ye2023feddisco} mitigate this problem. Momentum based stochastic weight averaging \cite{liu2025improving} and Iterative Moving Averaging (IMA) \cite{zhou2024understanding} further show that smoothing the model trajectory can improve generalization.

Under the above dual heterogeneity, unreliable pseudo labels corrupt localization model updates, while different client fingerprint distributions make the global localization model oscillate. We therefore propose FL-CsLoc-DH, a federated semi-supervised CSI crowdsensing framework for localization under dual heterogeneity. FL-CsLoc-DH combines three coordinated mechanisms: Two-Stage Noisy Gradient Rejection (TS-NGR), Soft Fisher Confidence Alignment (S-FCA), and Trajectory-Smoothed Model Averaging (T-SMA). At the server, TS-NGR estimates a robust loss consensus from participating clients and continuously downweights localization model updates whose degradation exceeds that consensus, thereby preserving useful information from heterogeneous regions instead of applying only a binary accept or reject rule. S-FCA contrasts labeled and unlabeled CSI fingerprint objective Fisher signals and applies a soft parameter mask to reduce unreliable localization model parameter updates. T-SMA applies an exponential moving average after a warmup phase to stabilize the global localization model trajectory. The main contributions of this work are summarized as follows:

\begin{itemize}
    \item We formulate the privacy-preserving FL-based crowdsensing localization problem under severe dual data heterogeneity. Specifically, we model the coupled challenges of spatial fingerprint feature shifts and highly skewed annotation ratios. Moreover, the problem treats heterogeneous measuring rates and partial participation as dynamic deployment conditions that intensify localization model deviation. The resulting formulation fully considers practical factors and guides the proposed algorithm toward real-world deployment.
    \item To address location label annotation heterogeneity, we propose a reliability control mechanism by TS-NGR. On the server, TS-NGR establishes a consensus based on the median and dynamically reweights clients exhibiting anomalous losses. Meanwhile, S-FCA applies a soft mask at the parameter level to limit the influence of updates with low confidence during local training. Together, these mechanisms stabilize the early training process and suppress confirmation bias without altering the core neural network architecture.
    \item To address data heterogeneity, we propose T-SMA. This mechanism applies an Exponential Moving Average (EMA) with dynamic momentum to the global localization model after aggregation. This design balances spatial exploration during early training with exploitation during later training, thereby ensuring smooth and monotonic convergence.
    \item We propose FL-CsLoc-DH by integrating TS-NGR, S-FCA, and T-SMA. Extensive experiments under isolated data or annotation heterogeneity and their dual setting demonstrate that FL-CsLoc-DH consistently improves localization accuracy and convergence stability. The ablation results further show that TS-NGR, S-FCA, and T-SMA provide complementary gains.
\end{itemize}

The remainder of this paper is organized as follows. Section II reviews related work. Section III introduces the crowdsensing localization system model and formulates dual heterogeneity. Section IV presents the proposed FL-CsLoc-DH framework. Section V reports the experimental evaluation, and Section VI concludes the paper.

\section{Related Work}

\subsection{Crowdsensing Localization}
MCS utilizes the mobile terminals of ordinary users for spatial fingerprint collection, significantly reducing the deployment cost of indoor localization. To optimize system performance, existing studies primarily focus on task allocation and data quality control. For instance, collaborative task allocation \cite{du2022collaborative} and coverage-aware strategies \cite{wang2024coverage} improve spatial sampling efficiency, while reputation-aware data fusion \cite{tang2018reputation} and reliability quantification models \cite{xie2021quantifying} are utilized to filter out low-quality crowdsourced contributions.

To leverage crowdsourced data while preserving user trajectory privacy, FL has been widely integrated into crowdsensing networks. Frameworks such as FedLoc \cite{yin2020fedloc}, CrowdFL \cite{zhao2022crowdfl}, and pseudo-label-driven FL-based crowdsourcing \cite{li2020pseudo} allow clients to retain raw Wi-Fi fingerprints locally and share only model parameter updates with the server. Although these methods demonstrate the feasibility of decentralized cooperative localization, they often struggle to handle the severe ``dual heterogeneity'' challenge in practical deployments. Channel feature shifts caused by physical environment differences and the extreme imbalance in client annotation ratios make the standard federated aggregation process highly susceptible to massive noisy pseudo-labels and unreliable localization model updates, ultimately causing the global localization model to deviate from the optimal convergence manifold.

\subsection{Federated Semi-Supervised Learning}
To address the widespread issues of label scarcity and pseudo-label noise in crowdsensing, FSSL provides an effective paradigm. By enabling decentralized clients to utilize massive unlabeled data alongside limited labeled samples, FSSL effectively alleviates the bottleneck of high annotation costs. To mitigate the confirmation bias caused by erroneous pseudo-labels, existing FSSL literature has introduced various robust mechanisms. Notable methods include establishing random sampling consensus \cite{liang2022rscfed}, applying class-balanced adaptive pseudo-labeling \cite{li2023class}, and selectively assimilating knowledge from limited labels \cite{cho2023local}. Other strategies involve multi-player pseudo-label generation networks \cite{che2021fedtrinet}, divergence learning for unlabeled clients \cite{elbatel2024learning}, and confidence discrepancy management \cite{liu2025mind}. Model robustness against scarce labels has also been studied in specific settings, such as $(FL)^2$ \cite{lee20242} and private semi-supervised learning \cite{fan2022private}. Meanwhile, centralized Semi-Supervised Learning (SSL) techniques, such as curriculum pseudo-labeling \cite{zhang2021flexmatch}, self-adaptive thresholding \cite{wang2022freematch}, and soft confidence weighting \cite{chen2023softmatch}, provide valuable insights for handling label noise. However, when client-specific label ratios and data streams vary unpredictably, it remains highly difficult to directly apply these SSL techniques, which are primarily designed for discrete classification tasks, to continuous CSI regression tasks.

\subsection{Dual Heterogeneity in Federated Semi-Supervised Learning}
More challengingly, in real-world crowdsensing deployments, the aforementioned imbalance in label ratios is often intertwined with physical environment differences, subjecting the system to a coupled dual heterogeneity with severe data distribution shifts across different physical environments (e.g., line-of-sight versus non-line-of-sight conditions) and extreme imbalances in annotation ratios across devices. Studies on cross-environment and heterogeneous databases consistently emphasize the significant impact of spatial variations on CSI fingerprints \cite{wang2015deepfi, wang2017cifi, chen2017confi,wang2018deep}. To tackle client drift caused by data heterogeneity, existing studies have proposed methods such as dynamic regularization \cite{acar2021federated}, model-contrastive alignment \cite{li2021model}, sharpness-aware minimization \cite{qu2022generalized}, and discrepancy-aware collaboration \cite{ye2023feddisco}. Furthermore, momentum-based stochastic weight averaging \cite{liu2025improving} and loss-landscape trajectory smoothing \cite{zhou2024understanding} demonstrate that late-stage averaging can effectively align solutions and improve model generalization under severe heterogeneity.

Specifically targeting this dual heterogeneity challenge, the FedF3 framework employs adaptive loss filtering and Fisher-based parameter selection to simultaneously handle data and label discrepancies \cite{peng2025navigating}. Nevertheless, because indoor localization inherently requires continuous coordinate regression, the hard rejection mechanisms used in methods like FedF3 may directly discard spatially rare yet highly valuable client updates. Therefore, to achieve high-precision indoor localization in extreme dual-heterogeneous environments, there is an urgent need to develop more adaptive parameter alignment and global trajectory smoothing strategies, ensuring that effective features from all types of clients are fully preserved and utilized.

\section{System Model}

\begin{figure}[!t]
    \centering
    \includegraphics[hiresbb,width=\columnwidth]{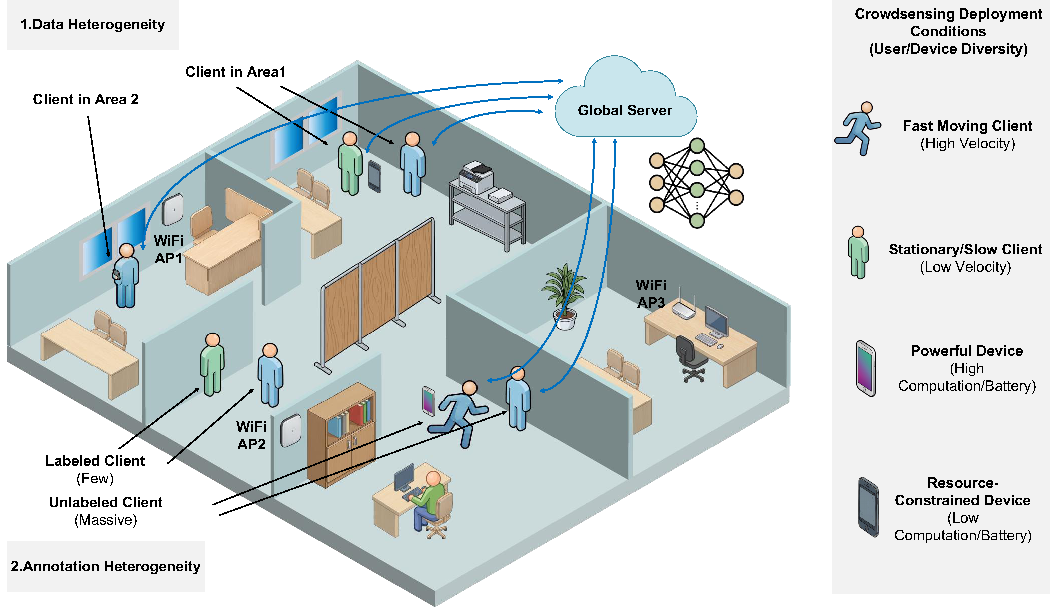}
    \caption{An FL-based crowdsensing indoor localization scenario under the dual heterogeneity. Clients in Areas 1 and 2 collect CSI from nearby APs in different propagation environments, producing data heterogeneity. Unequal availability of labeled and unlabeled fingerprints produces annotation heterogeneity. The right panel illustrates client mobility and differences in computation and battery capacity as deployment conditions.}
    \label{fig:environment}
\end{figure}

We consider the FL-based crowdsensing indoor localization scenario shown in Fig.~\ref{fig:environment}. Each mobile client continuously collects CSI fingerprints and measurement trajectories while moving through its own indoor region.

Each mobile client receives CSI from $A$ nearby Access Points (APs) over $M$ subcarriers. The complex CSI coefficient between AP $a$ and client $k$ on subcarrier $m$ is given by
\begin{equation}
H_{k,a,m}=\left|H_{k,a,m}\right|e^{\mathrm{j}\phi_{k,a,m}},
\label{eq:csi_coefficient}
\end{equation}
where $a\in\{1,\ldots,A\}$, $m\in\{1,\ldots,M\}$, $\left|H_{k,a,m}\right|$ is the amplitude, and $\phi_{k,a,m}$ is the phase. Here, $K$ is the total number of clients and $k\in\{1,\ldots,K\}$ indexes a client. The measurement index is omitted from $H_{k,a,m}$ for brevity. Following CSI fingerprint based localization \cite{wang2015deepfi,wang2017cifi,chen2017confi}, measurement $i$ at client $k$ is represented by
\begin{equation}
\begin{aligned}
\mathbf{x}_{k,i}=\bigl[&\left|H_{k,1,1}\right|,\ldots,\left|H_{k,A,M}\right|,\\
&\phi_{k,1,1},\ldots,\phi_{k,A,M}\bigr]^{\mathsf T}
\in\mathbb{R}^{2AM}.
\end{aligned}
\label{eq:csi_fingerprint}
\end{equation}

A fingerprint consists of CSI amplitude and phase. When the client reaches some reference positions, the simultaneously measured two-dimensional coordinates are attached to the fingerprints as their labels. Because ground-truth coordinates are unavailable at many positions and exhaustive manual calibration is costly, the remaining fingerprints are retained as unlabeled samples. Raw fingerprints and measurement trajectories remain on the collecting client throughout federated training.

The total number of communication rounds is $T$. The $K$ mobile clients and one central server jointly form a standard federated localization architecture \cite{yin2020fedloc,park2022federated,cheng2022federated}. During federated training, each client continues collecting fingerprints, so its local dataset at communication round $t$ is
\begin{equation}
\mathcal{D}_k^t \triangleq \mathcal{D}_k^{L,t}\cup\mathcal{D}_k^{U,t},
\end{equation}
where $\mathcal{D}_k^{L,t}\triangleq\{(\mathbf{x}_{k,i},\mathbf{y}_{k,i})\}_{i=1}^{n_{k,L}^t}$ contains fingerprints and the collecting positions and $\mathcal{D}_k^{U,t}\triangleq\{\mathbf{x}_{k,j}\}_{j=1}^{n_{k,U}^t}$ contains fingerprints without reliable ground truth. The vector $\mathbf{x}_{k,i}\in\mathbb{R}^{d}$ concatenates the $d$-dimensional CSI, i.e., the amplitude and phase features of the received signal, as specified in \eqref{eq:csi_fingerprint} with $d=2AM$, and $\mathbf{y}_{k,i}\in\mathbb{R}^{2}$ is the corresponding measurement coordinate. The quantities $n_{k,L}^t$ and $n_{k,U}^t$ denote the labeled and unlabeled sample counts, respectively. The total local sample count $n_k^t$ and label ratio $r_k^t$ are given by
\begin{equation}
n_k^t\triangleq n_{k,L}^t+n_{k,U}^t,\qquad
r_k^t\triangleq\frac{n_{k,L}^t}{n_k^t}.
\end{equation}

The local objective is constructed for continuous CSI coordinate regression by combining supervised empirical risk, a weighted pseudo label loss, and proximal regularization. The supervised term averages the Smooth L1 loss on measured coordinates; the unlabeled term uses model predictions as coordinate targets, following the use of unlabeled data in FSSL \cite{liang2022rscfed,peng2025navigating}. The proximal term follows FedProx \cite{li2020fedprox} and penalizes deviation from the broadcast parameters. Thus, the equation below combines these established learning and regularization principles in a local objective for continuous CSI coordinate regression.

The basic semi-supervised FL-based localization process follows four steps. First, the server selects the participating set $\mathcal{S}_t\subseteq\{1,\ldots,K\}$ and broadcasts the current global localization model $f_{\theta_{\mathrm{ima}}^{t-1}}$, where $\theta_{\mathrm{ima}}^{t-1}$ is its parameter vector before round $t$. Second, every selected client initializes its local localization model with $\theta_{\mathrm{ima}}^{t-1}$ and minimizes the semi-supervised local objective, which is expressed as
\begin{multline}
F_k^t(\theta)=\frac{1}{n_{k,L}^t}\sum_{i=1}^{n_{k,L}^t}\ell\bigl(f_{\theta}(\mathbf{x}_{k,i}),\mathbf{y}_{k,i}\bigr) \\
+\frac{\lambda_u^t}{n_{k,U}^t}\sum_{j=1}^{n_{k,U}^t}\ell\bigl(f_{\theta}(\mathcal{A}_t(\mathbf{x}_{k,j})),\widehat{\mathbf{y}}_{k,j}^{t}\bigr)
+\frac{\mu}{2}\lVert\theta-\theta_{\mathrm{ima}}^{t-1}\rVert_2^2,
\label{eq:local_fssl_objective}
\end{multline}
where $\theta$ is the trainable parameter vector, $f_{\theta}$ denotes the localization model parameterized by $\theta$, $\ell(\cdot)$ is the Smooth L1 localization loss, $\mathcal{A}_t$ is the stochastic training view, and $\widehat{\mathbf{y}}_{k,j}^{t}=\operatorname{sg}[f_{\theta}(\mathbf{x}_{k,j})]$ is the detached pseudo-label. The operator $\operatorname{sg}[\cdot]$ stops gradients, $\lambda_u^t\in[0,1]$ is the round-dependent unlabeled-loss weight, and $\mu\ge 0$ is the proximal coefficient. The unlabeled term is set to zero when $n_{k,U}^t=0$.

In local training, each unlabeled batch is first passed through the current localization model to obtain coordinate predictions, which are detached and used as fixed targets. The client then makes a separate stochastic forward pass and minimizes the weighted Smooth L1 loss between the new predictions and these targets. These operations are referred to as pseudo label training steps. Local optimization produces $\theta_{\mathrm{local},k}^{t}$.

Third, client $k$ uploads its refined localization model parameters $\theta_{\mathrm{upload}}^{t,(k)}$, its sample count $n_k^t$, and the summary statistic needed for reliability control, while keeping raw fingerprints and measurement coordinates local. Fourth, the server performs aggregation weighted by local sample counts, which is given by
\begin{equation}
\theta_{\mathrm{avg}}^t=\sum_{k\in\mathcal{S}_t}p_k^t\theta_{\mathrm{upload}}^{t,(k)},\qquad
p_k^t\triangleq\frac{n_k^t}{\sum_{q\in\mathcal{S}_t}n_q^t},
\label{eq:global_fssl_objective}
\end{equation}
where $\theta_{\mathrm{upload}}^{t,(k)}$ denotes the localization model parameter vector uploaded by client $k$ after local training in communication round $t$. In the basic process, $\theta_{\mathrm{upload}}^{t,(k)}=\theta_{\mathrm{local},k}^{t}$; in FL-CsLoc-DH, it is the S-FCA refined vector specified in \eqref{eq:sfca_uploaded_model}. The coefficient $p_k^t$ is the standard sample count weight and $\theta_{\mathrm{avg}}^t$ is the tentative global localization model parameter vector. After the server obtains the next broadcast state, these four steps repeat in round $t+1$.

The standard process is affected by two coupled forms of heterogeneity. Different clients collect fingerprints in different propagation environments, so their joint feature-coordinate distributions are non-identical. This produces data heterogeneity and drives local localization models toward different optima. Meanwhile, the availability of fingerprint labels varies across clients and rounds, so the location label annotation ratio $r_k^t$ differs substantially; this produces annotation heterogeneity and changes pseudo-label reliability. Continuous client movement and unequal collecting rates also change $n_k^t$ during training. These two forms of heterogeneity reduce the localization accuracy of the basic semi-supervised FL-based localization model and increase the localization error.

\section{Proposed FL-CsLoc-DH Framework}

\begin{figure}[!t]
    \centering
    \includegraphics[hiresbb,width=\columnwidth]{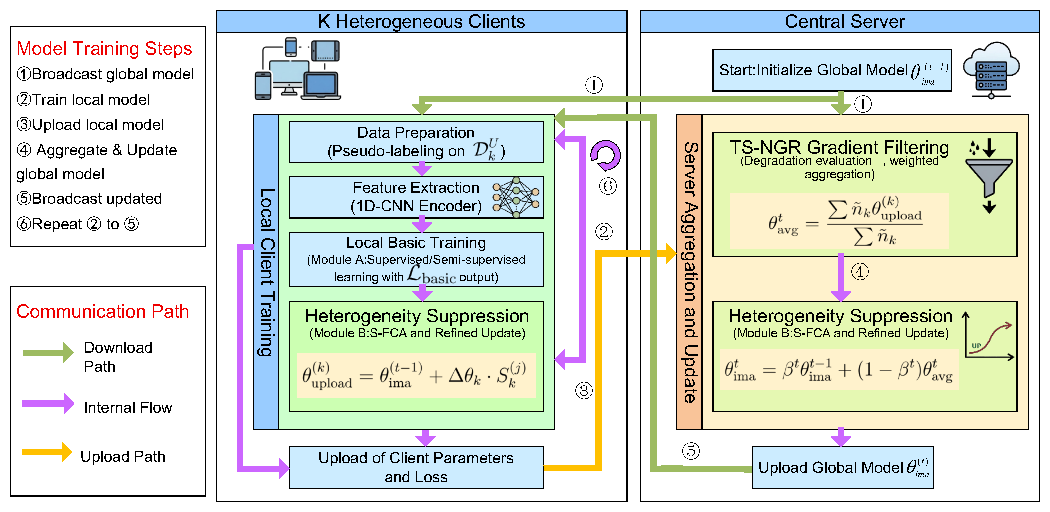}
    \caption{Overall workflow of FL-CsLoc-DH. Green arrows indicate server broadcasts, yellow arrows indicate client uploads, and purple arrows indicate internal processing. The 1D-CNN is a one-dimensional Convolutional Neural Network (CNN). Module A uses the local objective $F_k^t(\theta)$, labeled $\mathcal{L}_{\mathrm{basic}}$ in the diagram. Client-side S-FCA applies its mask componentwise to refine uploads, server-side TS-NGR weights client contributions, and the lower server averaging equation implements T-SMA. The bottom server block broadcasts the updated global model to clients. The round index $t$ is omitted from some upload and weighting symbols.}
    \label{fig:system}
\end{figure}

To achieve high-precision CSI indoor localization under the dual data heterogeneity, this paper proposes FL-CsLoc-DH. As illustrated in Fig.~\ref{fig:system}, the framework coordinates three components along the federated update path. TS-NGR controls client-level reliability at the server, S-FCA controls parameter-level reliability during local training, and T-SMA stabilizes the global localization model trajectory after aggregation.

\subsection{Two-Stage Noisy Gradient Rejection (TS-NGR)}

Under the dual data heterogeneity, the conventional Federated Averaging (FedAvg) algorithm \cite{mcmahan2017communication} allocates aggregation weights solely based on the data volume of each client. However, due to the superposition of label skewness and differences in wireless propagation in the collecting area, certain clients may yield extraordinarily high localization errors, transmitting localization model parameters heavily corrupted by noise. To defend against these degraded updates, we propose the TS-NGR mechanism based on a performance degradation penalty at the server side.

Consistent with the implementation, client $k$ reports the mean supervised local objective function accumulated over the $B_k^t$ labeled mini-batches processed in round $t$. The objective function is given by
\begin{equation}
L_k^t=\frac{1}{B_k^t}\sum_{b=1}^{B_k^t}\left[\ell_{k,b}^{L,t}
+\frac{\mu}{2}\lVert\theta_{k,b}^t-\theta_{\mathrm{ima}}^{t-1}\rVert_2^2\right],
\label{eq:reported_local_loss}
\end{equation}
where $B_k^t$ is the number of labeled mini-batches processed by client $k$ in round $t$, $\ell_{k,b}^{L,t}$ is the mean labeled Smooth L1 loss of mini-batch $b$, and $\theta_{k,b}^t$ is the parameter vector of the corresponding local localization model. Pseudo label training steps affect the localization model update and Fisher statistics but are not counted again in $L_k^t$. Thus, $L_k^t$ summarizes the labeled-loss and proximal terms of $F_k^t(\theta)$ in \eqref{eq:local_fssl_objective}, rather than redefining the local objective function. Each client uploads only $L_k^t$, $n_k^t$, and its S-FCA-refined localization model $\theta_{\mathrm{upload}}^{t,(k)}$.

In round $t$, the server receives $L_k^t$ from every client $k\in\mathcal{S}_t$. It first computes the median loss as follows
\begin{equation}
    L_{\mathrm{med}}^t = \mathrm{Median}(\{L_k^t \mid k \in \mathcal{S}_t\}).
\end{equation}
Compared with the mean, the median is more robust to anomalous losses. The performance degradation ratio of client $k$ is given by
\begin{equation}
    d_k^t = \frac{\max(0,L_k^t-L_{\mathrm{med}}^t)}{L_{\mathrm{med}}^t+\epsilon},
\end{equation}
where $\epsilon$ is a small positive constant that prevents division by zero. A larger $d_k^t$ indicates a less reliable client update.

TS-NGR converts this ratio into a penalty $a_k^t$ and an adaptive client mask $m_k^t$, given by
\begin{equation}
    a_k^t=\eta d_k^t,
\end{equation}
\begin{equation}
    m_k^t =
    \begin{cases}
    \omega_{\mathrm{min}}, & d_k^t>\tau_{\mathrm{deg}}, \\
    \max(\omega_{\mathrm{base}},\exp(-a_k^t)), & \text{otherwise}.
    \end{cases}
\end{equation}
The coefficient $\eta$ controls the decay rate and $\tau_{\mathrm{deg}}$ is the extreme anomaly threshold. The floor values $\omega_{\mathrm{min}}$ and $\omega_{\mathrm{base}}$ prevent complete removal of spatially rare information.

The effective aggregation sample size is given by
\begin{equation}
    \widetilde{n}_k^t=n_k^t m_k^t.
\end{equation}
The server therefore retains every selected client while continuously reducing the influence of degraded updates.

\subsection{Soft Fisher Confidence Alignment (S-FCA)}

While the server-side degradation penalty filters low-quality clients at the macroscopic level, it is imperative to align features at the microscopic parameter level during local training. To this end, we propose the S-FCA mechanism. S-FCA evaluates the reliability of individual localization model parameters by comparing their sensitivities to labeled versus unlabeled data.

During local optimization, client $k$ approximates the diagonal Fisher information for parameter index $j$ using squared gradients from the labeled and unlabeled steps, which is written as
\begin{equation}
    F_{L,k}^{t,(j)}=\sum_{(\mathbf{x},\mathbf{y})\in\mathcal{D}_k^{L,t}}
    \left(\nabla_{\theta_j}\ell(f_{\theta}(\mathbf{x}),\mathbf{y})\right)^2,
\end{equation}
\begin{equation}
    F_{U,k}^{t,(j)}=\sum_{\mathbf{x}\in\mathcal{D}_k^{U,t}}
    \left(\nabla_{\theta_j}\ell(f_{\theta}(\mathcal{A}_t(\mathbf{x})),\widehat{\mathbf{y}}^t)\right)^2.
\end{equation}
The quantities $F_{L,k}^{t,(j)}$ and $F_{U,k}^{t,(j)}$ denote the accumulated squared-gradient estimates of the diagonal Fisher information for localization model parameter $j$ from labeled and unlabeled steps, respectively. In the unlabeled expression, $\widehat{\mathbf{y}}^t$ abbreviates the detached pseudo-label $\widehat{\mathbf{y}}_{k,j}^{t}$ associated with the current sample, as defined in \eqref{eq:local_fssl_objective}; the parameter index $j$ in the Fisher statistic is independent of the sample index. Parameters dominated by $F_{U,k}^{t,(j)}$ are more susceptible to pseudo-label noise. Their composite importance is defined as
\begin{equation}
    I_k^{t,(j)}=\gamma_F F_{L,k}^{t,(j)}-\lambda_F F_{U,k}^{t,(j)},
\end{equation}
where $\gamma_F$ and $\lambda_F$ are prespecified Fisher trade-off constants.

The pseudo-label weight $\lambda_u^t$ is the same coefficient as in \eqref{eq:local_fssl_objective}, not a new parameter.

Let $P$ denote the number of localization model parameters and let $\alpha_{\mathrm{ffs}}\in(0,1]$ be the prespecified fraction retained in full. Sort the scores as $I_k^{t,[1]}\geq\cdots\geq I_k^{t,[P]}$ and set $h=\max(1,\lfloor\alpha_{\mathrm{ffs}}P\rfloor)$. The threshold selecting the prescribed top fraction is given by
\begin{equation}
\tau_{k,\mathrm{ffs}}^t=I_k^{t,[h]}.
\label{eq:ffs_threshold}
\end{equation} S-FCA forms the hard indicator and its soft counterpart as
\begin{equation}
    M_k^{t,(j)}=\mathbb{I}\left(I_k^{t,(j)}\ge\tau_{k,\mathrm{ffs}}^t\right),
\end{equation}
\begin{equation}
    S_k^{t,(j)}=M_k^{t,(j)}+\left(1-M_k^{t,(j)}\right)\rho,
\end{equation}
where $\mathbb{I}(\cdot)$ is the indicator function, equal to 1 when its condition holds and 0 otherwise, and $\rho\in(0,1)$ is the prespecified leakage rate. Reliable parameters retain their full increments, whereas the remaining parameters retain the fraction $\rho$.

Using the broadcast localization model $\theta_{\mathrm{ima}}^{t-1}$, the refined increment and upload are given by
\begin{equation}
    \Delta\theta_k^{t,(j)}=\theta_{\mathrm{local},k}^{t,(j)}-
    (\theta_{\mathrm{ima}}^{t-1})^{(j)},
\end{equation}
\begin{equation}
    \theta_{\mathrm{upload},j}^{t,(k)}=(\theta_{\mathrm{ima}}^{t-1})^{(j)}
    +S_k^{t,(j)}\Delta\theta_k^{t,(j)}.
    \label{eq:sfca_uploaded_model}
\end{equation}

\subsection{Trajectory-Smoothed Model Averaging (T-SMA)}

Although S-FCA and TS-NGR mitigate annotation skewness, intrinsic spatial data heterogeneity continues to compel local localization models toward divergent optima. To fundamentally resolve the trajectory divergence without imposing additional memory burdens, we propose the T-SMA mechanism. T-SMA orchestrates an EMA fortified by a time-based dynamic momentum scheduler. This design is motivated by model trajectory averaging \cite{zhou2024understanding} and momentum-based weight averaging \cite{liu2025improving}.

Using the TS-NGR effective sample sizes, the tentative aggregate at the server in round $t$ is expressed as
\begin{equation}
    \theta_{\mathrm{avg}}^t=\sum_{k\in\mathcal{S}_t}
    \frac{\widetilde{n}_k^t}{\sum_{q\in\mathcal{S}_t}\widetilde{n}_q^t}
    \theta_{\mathrm{upload}}^{t,(k)}.
\end{equation}
After the activation round, defined as $T_{\mathrm{start}}$, normalized training progress is given by
\begin{equation}
    \pi(t)=\min\left(1,\max\left(0,
    \frac{t-T_{\mathrm{start}}}{T-T_{\mathrm{start}}}\right)\right),
\end{equation}
\begin{equation}
    \beta^t=\beta_{\mathrm{base}}+\Delta\beta\,\pi(t).
\end{equation}
Here, $\beta_{\mathrm{base}}$ is the momentum at activation and $\Delta\beta$ is its prescribed increase over the remaining rounds, with $0\leq\beta_{\mathrm{base}}\leq\beta_{\mathrm{base}}+\Delta\beta<1$. The activation round satisfies $0\leq T_{\mathrm{start}}<T$.
The localization model broadcast in the next round is given by
\begin{equation}
    \theta_{\mathrm{ima}}^t=\beta^t\theta_{\mathrm{ima}}^{t-1}
    +(1-\beta^t)\theta_{\mathrm{avg}}^t.
\end{equation}
Before $T_{\mathrm{start}}$, we set $\theta_{\mathrm{ima}}^t=\theta_{\mathrm{avg}}^t$. Thus, early rounds rapidly absorb new propagation features, whereas the increasing $\beta^t$ suppresses late-stage trajectory oscillations. The subscript ``ima'' identifies the persistent averaging state and does not denote the ordinary IMA baseline \cite{zhou2024understanding}.

\begin{algorithm}[t]
\caption{FL-CsLoc-DH Server-Side Execution}
\label{alg:server}
\begin{algorithmic}[1]
\Require Initial localization model $\theta_{\mathrm{ima}}^0$, rounds $T$, hyperparameters $\epsilon,\eta,\tau_{\mathrm{deg}},\omega_{\mathrm{min}},\omega_{\mathrm{base}},\beta_{\mathrm{base}},\Delta\beta,T_{\mathrm{start}}$
\Ensure Final global localization model $\theta_{\mathrm{ima}}^{T}$
\For{$t=1,\dots,T$}
    \State Select $\mathcal{S}_t$ and broadcast $\theta_{\mathrm{ima}}^{t-1}$

    \State \textbf{Step 1: Local Client Updates (Parallel Execution)}
    \For{client $k \in \mathcal{S}_t$ \textbf{in parallel}}
        \State $(L_k^t,\theta_{\mathrm{upload}}^{t,(k)},n_k^t)\leftarrow\text{ClientUpdate}(k,t,\theta_{\mathrm{ima}}^{t-1})$
    \EndFor

    \State \textbf{Step 2: Two-Stage Noisy Gradient Rejection (TS-NGR)}
    \State $L_{\mathrm{med}}^t\leftarrow\mathrm{Median}(\{L_k^t\mid k\in\mathcal{S}_t\})$
    \For{client $k \in \mathcal{S}_t$}
        \State $d_k^t\leftarrow\frac{\max(0,L_k^t-L_{\mathrm{med}}^t)}{L_{\mathrm{med}}^t+\epsilon}$; $a_k^t\leftarrow\eta d_k^t$
        \If{$d_k^t>\tau_{\mathrm{deg}}$}
            \State $m_k^t\leftarrow\omega_{\mathrm{min}}$
        \Else
            \State $m_k^t\leftarrow\max(\omega_{\mathrm{base}},\exp(-a_k^t))$
        \EndIf
        \State $\widetilde{n}_k^t\leftarrow n_k^t m_k^t$
    \EndFor

    \State \textbf{Step 3: Trajectory-Smoothed Model Averaging (T-SMA)}
    \State $\theta_{\mathrm{avg}}^t\leftarrow\sum_{k\in\mathcal{S}_t}\frac{\widetilde{n}_k^t}{\sum_{q\in\mathcal{S}_t}\widetilde{n}_q^t}\theta_{\mathrm{upload}}^{t,(k)}$
    \If{$t<T_{\mathrm{start}}$}
        \State $\theta_{\mathrm{ima}}^t\leftarrow\theta_{\mathrm{avg}}^t$
    \Else
        \State $\pi(t)\leftarrow\min\left(1,\max\left(0,\frac{t-T_{\mathrm{start}}}{T-T_{\mathrm{start}}}\right)\right)$
        \State $\beta^t\leftarrow\beta_{\mathrm{base}}+\Delta\beta\,\pi(t)$
        \State $\theta_{\mathrm{ima}}^t\leftarrow\beta^t\theta_{\mathrm{ima}}^{t-1}+(1-\beta^t)\theta_{\mathrm{avg}}^t$
    \EndIf
\EndFor
\State \Return $\theta_{\mathrm{ima}}^{T}$
\end{algorithmic}
\end{algorithm}

\begin{algorithm}[t]
\caption{Local Client Update}
\label{alg:client}
\begin{algorithmic}[1]
\Require Client $k$, round $t$, localization model $\theta_{\mathrm{ima}}^{t-1}$, data $\mathcal{D}_k^{L,t},\mathcal{D}_k^{U,t}$, epochs $E$, coefficients $\mu,\lambda_u^t,\rho,\gamma_F,\lambda_F,\alpha_{\mathrm{ffs}}$
\Ensure Statistic $L_k^t$, upload $\theta_{\mathrm{upload}}^{t,(k)}$, sample count $n_k^t$
\State $\theta_{\mathrm{local},k}^t\leftarrow\theta_{\mathrm{ima}}^{t-1}$; initialize Fisher accumulators and loss list $\mathcal{B}_k^t$

\State \textbf{Step 1: Local Optimization}
\For{$e=1,\dots,E$}
    \State Optimize labeled batches using $\ell^L+\frac{\mu}{2}\lVert\theta_{\mathrm{local},k}^t-\theta_{\mathrm{ima}}^{t-1}\rVert_2^2$
    \State Append each labeled batch objective to $\mathcal{B}_k^t$ and accumulate $F_{L,k}^{t,(j)}$
    \State Generate detached $\widehat{\mathbf{y}}^t$ and optimize unlabeled batches using $\lambda_u^t\ell^U$
    \State Accumulate $F_{U,k}^{t,(j)}$
\EndFor
\State $L_k^t\leftarrow|\mathcal{B}_k^t|^{-1}\sum_{z\in\mathcal{B}_k^t}z$

\State \textbf{Step 2: Parameter-Level S-FCA}
\For{each parameter index $j$}
    \State $I_k^{t,(j)}\leftarrow\gamma_F F_{L,k}^{t,(j)}-\lambda_F F_{U,k}^{t,(j)}$
\EndFor
\State Determine $\tau_{k,\mathrm{ffs}}^t$ from the top $\alpha_{\mathrm{ffs}}$ fraction of $\{I_k^{t,(j)}\}$

\State \textbf{Step 3: Soft Masking and Weight Refinement}
\For{each parameter index $j$}
    \State $M_k^{t,(j)}\leftarrow\mathbb{I}(I_k^{t,(j)}\ge\tau_{k,\mathrm{ffs}}^t)$
    \State $S_k^{t,(j)}\leftarrow M_k^{t,(j)}+(1-M_k^{t,(j)})\rho$
    \State $\Delta\theta_k^{t,(j)}\leftarrow\theta_{\mathrm{local},k}^{t,(j)}-(\theta_{\mathrm{ima}}^{t-1})^{(j)}$
    \State $\theta_{\mathrm{upload},j}^{t,(k)}\leftarrow(\theta_{\mathrm{ima}}^{t-1})^{(j)}+S_k^{t,(j)}\Delta\theta_k^{t,(j)}$
\EndFor
\State $n_k^t\leftarrow n_{k,L}^t+n_{k,U}^t$; \Return $L_k^t,\theta_{\mathrm{upload}}^{t,(k)},n_k^t$
\end{algorithmic}
\end{algorithm}

\subsection{Overall FL-CsLoc-DH Procedure}

Algorithm~\ref{alg:server} summarizes the server operations in each communication round. It invokes the client procedure, computes the TS-NGR weights, aggregates the refined uploads, and applies T-SMA. Algorithm~\ref{alg:client} specifies the local optimization and S-FCA refinement performed by every selected client. In Algorithm~\ref{alg:client}, $\ell^L$ and $\ell^U$ denote the labeled and unlabeled batch-loss terms of $F_k^t(\theta)$, respectively.

As shown in Fig.~\ref{fig:system}, FL-CsLoc-DH follows the same client training, upload, server aggregation, and rebroadcast path as the basic semi-supervised FL localization process. At round $t$, the server broadcasts $\theta_{\mathrm{ima}}^{t-1}$ to the selected clients. Each client uses the 1D-CNN encoder and performs the local objective $F_k^t(\theta)$, labeled $\mathcal{L}_{\mathrm{basic}}$ in Fig.~\ref{fig:system}, computes the two Fisher statistics, and uses S-FCA to refine its localization model increment as detailed in Algorithm~\ref{alg:client}. The server receives only the refined model, the reported labeled objective, and the sample count. The server, following Algorithm~\ref{alg:server}, then uses TS-NGR to convert the reported objectives into continuous client weights, forms $\theta_{\mathrm{avg}}^t$, and applies T-SMA after $T_{\mathrm{start}}$. The resulting $\theta_{\mathrm{ima}}^t$ is broadcast in the next round. This procedure coordinates parameter reliability, client reliability, and global localization model parameter stability without uploading raw CSI fingerprints or coordinates.

\section{Experimental Evaluation}
\label{sec:experiments}

This section evaluates FL-CsLoc-DH in the crowdsensing localization scenario. We first describe the data, implementation settings, baselines, and metrics. We then compare the proposed method with FedAvg \cite{mcmahan2017communication}, FedF3 \cite{peng2025navigating}, and IMA \cite{zhou2024understanding}, examine the complete dual heterogeneity setting, quantify the contribution of each component, and analyze the T-SMA momentum parameter. The experiments address three questions: whether FL-CsLoc-DH improves localization under dual heterogeneity, whether TS-NGR, S-FCA, and T-SMA provide complementary gains, and whether these gains are obtained with practical convergence time.

\subsection{Experimental Setup}

\subsubsection{Dataset and preprocessing}
The main comparison uses 1,200 CSI fingerprints selected from the open-access ultra dense indoor MaMIMO CSI dataset described in \cite{li2022toward}. The dataset was measured in an indoor laboratory with 64 distributed antennas, 100 subcarriers, and user positions controlled by a computer numerical control positioning table. Each complex CSI measurement is represented by concatenating the amplitude and phase information of the measuring signal, and the prediction target is the two-dimensional client coordinate. We randomly allocate 80\% of the samples to training and 20\% to testing using a fixed split seed of 42. The CSI features are standardized using statistics computed only from the training partition. Coordinate targets are scaled to $[-1,1]$ during optimization and are transformed back to physical units for localization error evaluation.

The six spatial regions form six clients, which creates data heterogeneity because each client observes a physically isolated CSI distribution. Annotation heterogeneity is imposed through client label ratios of 80\%, 50\%, 30\%, 10\%, 5\%, and 1\%. The remaining samples are treated as unlabeled data and participate through pseudo-label training. Three clients are sampled in every communication round. This protocol directly couples spatial data shifts with strongly imbalanced annotation availability.

For the controlled ablation and sensitivity studies, we additionally use three spatial clients corresponding to the northeast, northwest, and southwest regions of the ultra-dense CSI data. Each client contains 170 samples and uses label ratios of 80\%, 10\%, and 2\%, respectively. Moderate, severe, and extreme spatial heterogeneity are constructed with regional sampling intervals of $5/20/30$ mm, $5/30/50$ mm, and $5/40/80$ mm. During training only, transient pseudo-label perturbations with standard deviations of 0.12, 0.16, and 0.20 are applied to the sparsest region for these three settings. Validation and test samples remain unperturbed.

\subsubsection{Localization model and optimization}
All methods use the same lightweight one-dimensional CNN. The network contains two convolutional layers with 16 and 32 channels, max pooling, a 128-unit fully connected layer, dropout of 0.4, and a two-dimensional regression head. Unless otherwise stated, training lasts 400 communication rounds, each selected client performs three local epochs with batch size 32, and Adam is used with an initial learning rate of $10^{-3}$ and weight decay of $10^{-4}$. The learning rate decays once per communication round. A proximal coefficient of 0.05 is used for the semi-supervised federated variants. The pseudo-label warm-up lasts 40 rounds in the controlled experiments.

The full three-module ablation and the $\beta_{\mathrm{base}}$ sweep are single-seed controlled studies, so their numerical differences are interpreted as diagnostic evidence rather than confidence intervals.

\subsubsection{Baselines and metrics}
We compare against centralized training, FedAvg \cite{mcmahan2017communication}, a FedF3-inspired implementation with hard client filtering and hard Fisher parameter selection \cite{peng2025navigating}, and ordinary IMA with a five-round sliding window \cite{zhou2024understanding}. Here, the FedAvg baseline uses sample-count-weighted aggregation with semi-supervised local training in Figs.~\ref{fig:fedf3_uniform} and~\ref{fig:dual_comparison}, and supervised local training in Fig.~\ref{fig:ima_comparison}. The IMA and dual-heterogeneity comparisons retain proximal regularization \cite{li2020fedprox}, whereas the FedF3 comparison uses no proximal term. Centralized training is included only as a non-private upper reference. All federated methods use the same backbone and communication budget. Performance is measured by test Smooth L1 loss in the normalized coordinate space and mean Euclidean localization error after inverse scaling. Lower values are better. Convergence time combines the communication delay and local computation assumed by the timing experiment.

\subsection{Performance under Isolated Heterogeneity}

\begin{figure}[!t]
    \centering
    \includegraphics[hiresbb,width=\columnwidth]{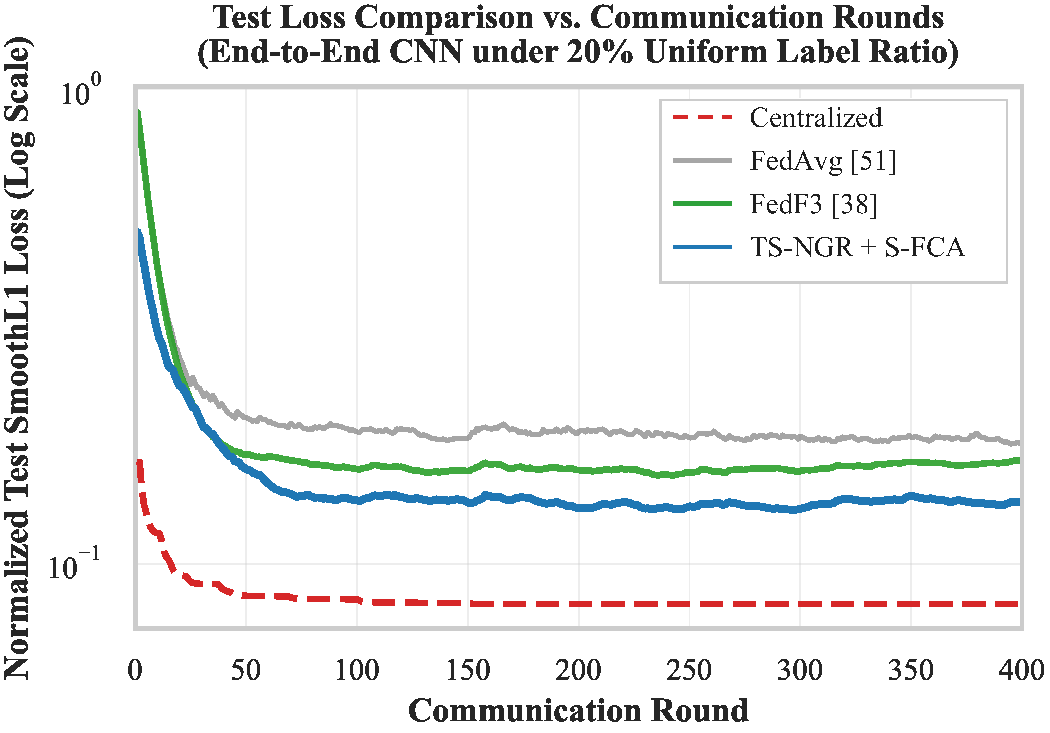}
    \caption{Dimensionless normalized-coordinate test Smooth L1 loss comparison of TS-NGR with S-FCA against the FedF3 baseline under a uniform 20\% client label ratio.}
    \label{fig:fedf3_uniform}
\end{figure}

Fig.~\ref{fig:fedf3_uniform} shows test Smooth L1 loss curves under data heterogeneity with a uniform 20\% label ratio at every client. Thus, annotation heterogeneity is removed while data heterogeneity is retained. At the final displayed round, TS-NGR with S-FCA reaches a smoothed test loss of approximately 0.1351, compared with 0.1651 for the FedF3 baseline \cite{peng2025navigating} and 0.1792 for the semi-supervised FedAvg baseline \cite{mcmahan2017communication}. This corresponds to reductions of about 18.2\% and 24.6\%, respectively. The result indicates that soft parameter retention and soft client reweighting preserve useful regional information that can be removed by hard rejection.

\begin{figure}[!t]
    \centering
    \includegraphics[hiresbb,width=\columnwidth]{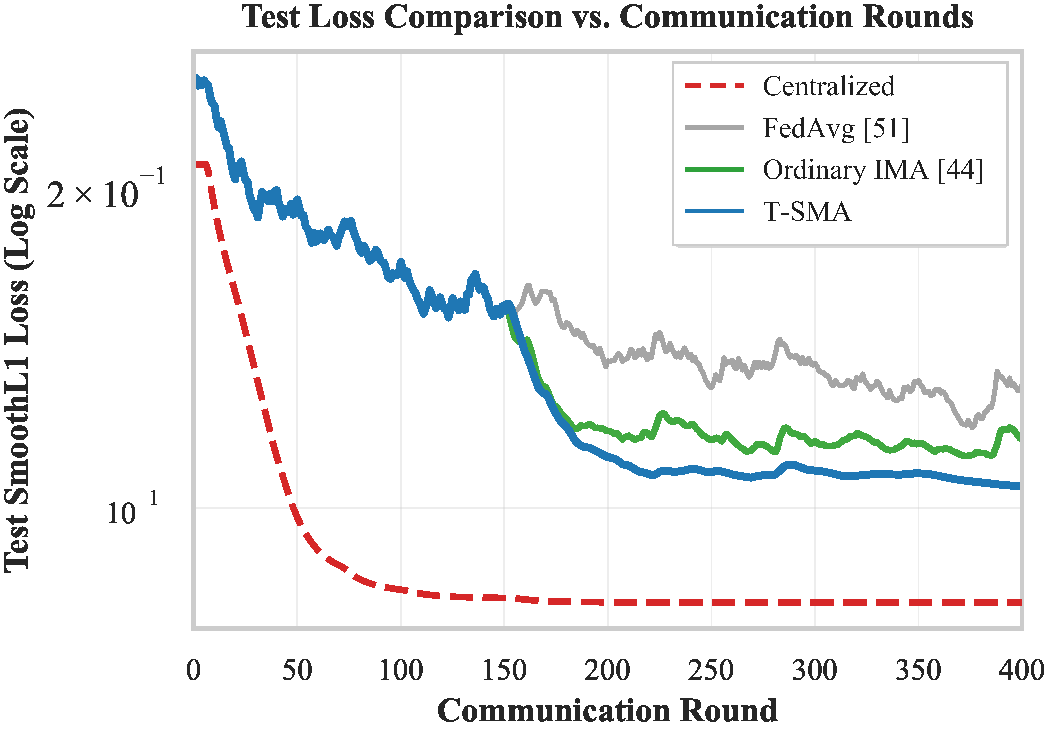}
    \caption{Test Smooth L1 loss comparison between T-SMA, ordinary IMA, and FedAvg under spatially heterogeneous CSI data.}
    \label{fig:ima_comparison}
\end{figure}

Fig.~\ref{fig:ima_comparison} compares test Smooth L1 loss of dynamic T-SMA and ordinary IMA \cite{zhou2024understanding} under data heterogeneity. All samples are labeled in this comparison, so annotation heterogeneity is absent. This comparison uses five local epochs and one selected client per round. Both averaging schemes start after the warm-up stage in this controlled experiment. T-SMA reaches approximately 0.1049 test loss, whereas ordinary IMA and the supervised FedAvg baseline \cite{mcmahan2017communication} reach approximately 0.1163 and 0.1311. The respective reductions are 9.8\% and 20.0\%. The increasing momentum used by T-SMA therefore suppresses late-round oscillation more effectively than a fixed sliding window while still allowing earlier global updates to follow newly observed spatial features.

\subsection{Performance under Dual Heterogeneity}

\begin{figure}[!t]
    \centering
    \includegraphics[hiresbb,width=\columnwidth]{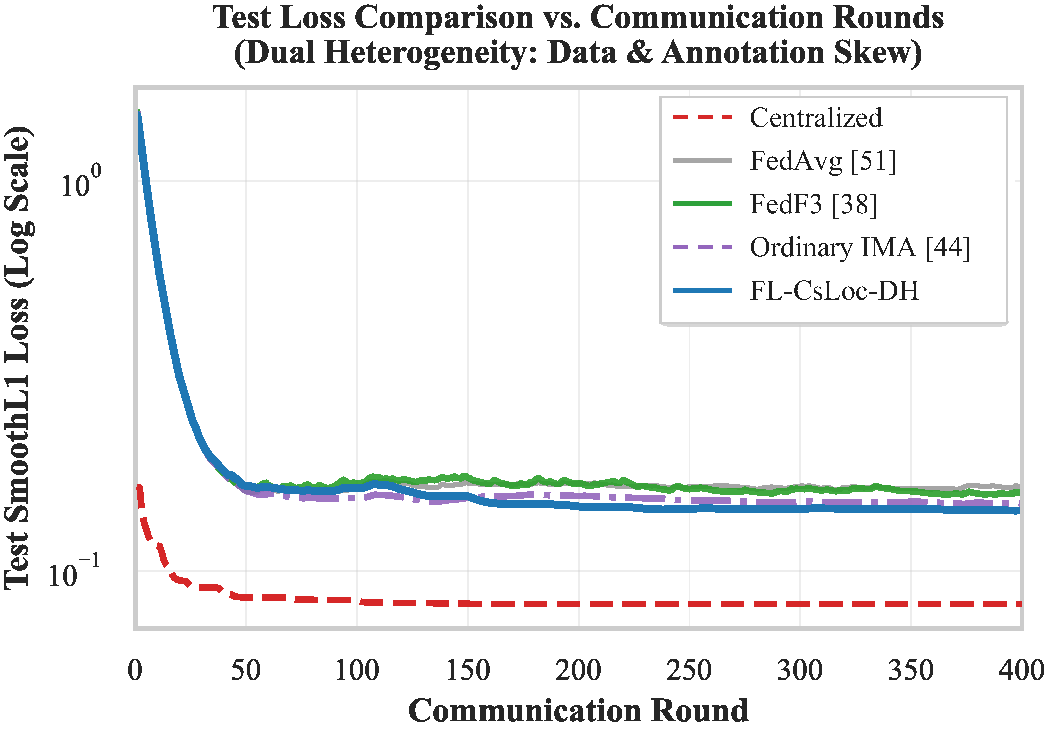}
    \caption{Test Smooth L1 loss comparison with FedF3 and ordinary IMA under dual spatial and annotation heterogeneity. FL-CsLoc-DH combines TS-NGR, S-FCA, and T-SMA.}
    \label{fig:dual_comparison}
\end{figure}

Fig.~\ref{fig:dual_comparison} presents the main comparison in which the six spatial clients also have label ratios ranging from 80\% to 1\%. The full method has the lowest test loss throughout the late training stage and exhibits the smoothest global localization model test-loss trajectory after its reliability and averaging mechanisms are activated. At round 400, the displayed losses of the semi-supervised FedAvg baseline \cite{mcmahan2017communication}, FedF3, ordinary IMA, and FL-CsLoc-DH are approximately 0.1636, 0.1586, 0.1491, and 0.1428. FL-CsLoc-DH therefore reduces loss by 12.7\% relative to FedAvg, 10.0\% relative to FedF3, and 4.2\% relative to ordinary IMA.

The remaining gap to centralized training is expected because the centralized reference observes all raw labeled samples without privacy or distribution constraints. Nevertheless, the lower loss relative to both FedF3 and IMA shows that client reliability, parameter reliability, and trajectory smoothing address different parts of the dual heterogeneity problem.

\subsection{Ablation Study}

Table~\ref{tab:module_ablation} reports a controlled ablation of the three proposed modules. The full method achieves the lowest loss and mean error in all three tested settings. Relative to semi-supervised FedAvg, it reduces loss by 21.2\%, 15.4\%, and 8.9\% for moderate, severe, and extreme heterogeneity. The corresponding error reductions are 13.6\%, 8.8\%, and 6.4\%. Removing TS-NGR causes the smallest degradation in the moderate setting but a clear loss increase in the severe and extreme settings. Removing S-FCA yields the largest deterioration in most columns, while removing T-SMA consistently reduces late-stage accuracy. These results support the intended division of labor among robust client weighting, parameter confidence alignment, and global trajectory smoothing.

\begin{table*}[!t]
\centering
\scriptsize
\caption{Final-round module ablation under dual heterogeneity. Sampling intervals are listed in northeast, northwest, and southwest order.}
\label{tab:module_ablation}
\resizebox{\textwidth}{!}{%
\begin{tabular}{lccccccccc}
\toprule
Experiment & TS-NGR & S-FCA & T-SMA & \multicolumn{2}{c}{Moderate, 5/20/30 mm} & \multicolumn{2}{c}{Severe, 5/30/50 mm} & \multicolumn{2}{c}{Extreme, 5/40/80 mm} \\
\cmidrule(lr){5-6}\cmidrule(lr){7-8}\cmidrule(lr){9-10}
 & & & & Loss & Error (m) & Loss & Error (m) & Loss & Error (m) \\
\midrule
Semi-supervised FedAvg & $\times$ & $\times$ & $\times$ & 0.1523 & 1.03611 & 0.1705 & 1.08639 & 0.1546 & 1.03318 \\
w/o TS-NGR & $\times$ & $\checkmark$ & $\checkmark$ & 0.1205 & 0.89759 & 0.1455 & 1.00148 & 0.1412 & 0.97305 \\
w/o S-FCA & $\checkmark$ & $\times$ & $\checkmark$ & 0.1291 & 0.92860 & 0.1526 & 1.04415 & 0.1514 & 1.03451 \\
w/o T-SMA & $\checkmark$ & $\checkmark$ & $\times$ & 0.1254 & 0.93671 & 0.1505 & 1.00572 & 0.1447 & 0.98772 \\
\textbf{FL-CsLoc-DH} & $\checkmark$ & $\checkmark$ & $\checkmark$ & \textbf{0.1201} & \textbf{0.89570} & \textbf{0.1443} & \textbf{0.99116} & \textbf{0.1408} & \textbf{0.96702} \\
\bottomrule
\end{tabular}}
\end{table*}

\begin{figure*}[!t]
    \centering
    \begin{minipage}{0.49\textwidth}
        \centering
        \includegraphics[hiresbb,width=\linewidth]{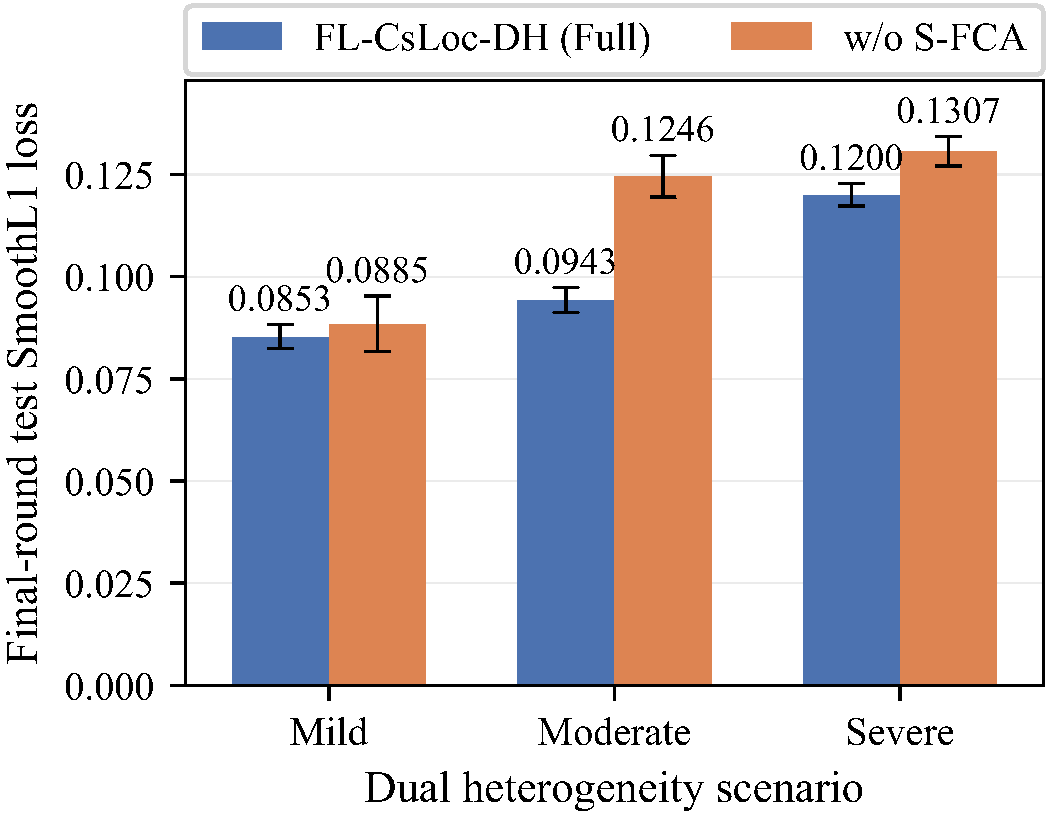}\\[-1mm]
        \small (a) Final-round Smooth L1 loss
    \end{minipage}
    \hfill
    \begin{minipage}{0.49\textwidth}
        \centering
        \includegraphics[hiresbb,width=\linewidth]{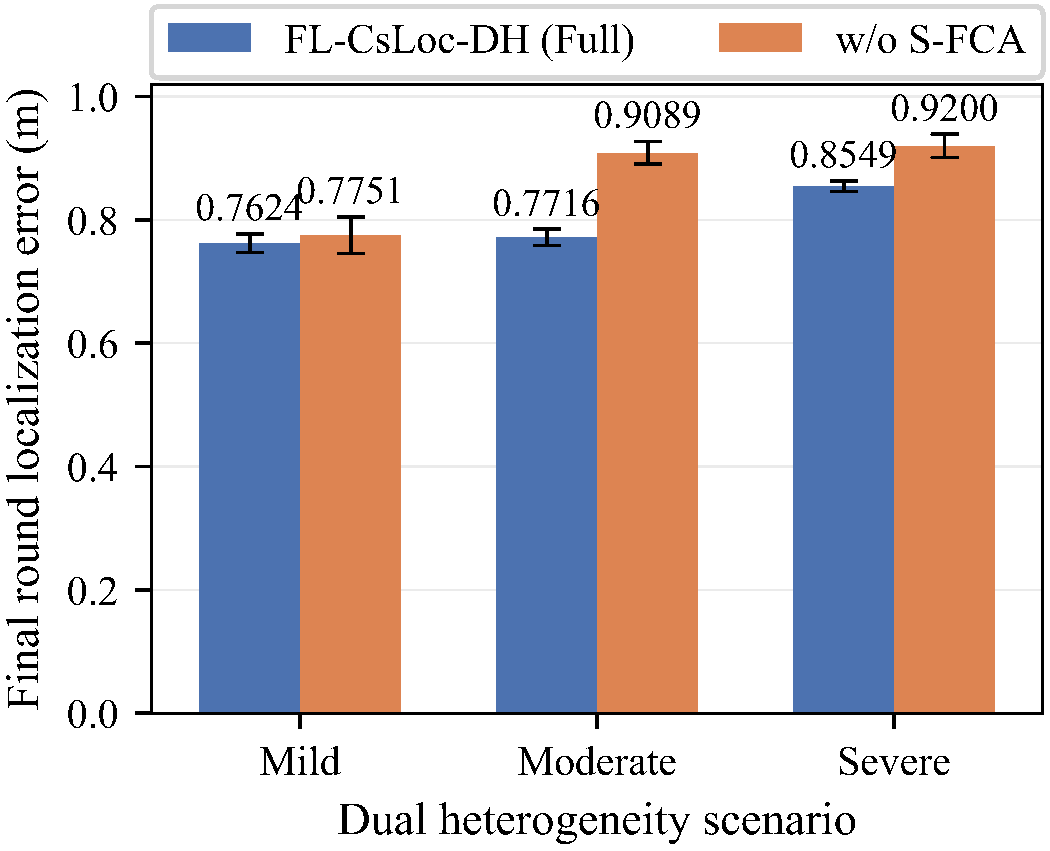}\\[-1mm]
        \small (b) Final-round localization error
    \end{minipage}
    \caption{S-FCA ablation across dual heterogeneity levels. Bars show the mean over three seeds and error bars show one sample standard deviation.}
    \label{fig:s_fca_ablation}
\end{figure*}

The focused S-FCA study in Fig.~\ref{fig:s_fca_ablation} provides three-seed evidence for the parameter-level mechanism. Disabling S-FCA increases mean test loss from 0.0853 to 0.0885 in the mild setting, from 0.0943 to 0.1246 in the moderate setting, and from 0.1200 to 0.1307 in the severe setting. The corresponding mean errors increase from 0.7624 to 0.7751 m, from 0.7716 to 0.9089 m, and from 0.8549 to 0.9200 m. The largest loss and error reductions, 24.3\% and 15.1\%, occur in the moderate setting. In the mild setting, the smaller difference and overlapping error bars indicate that aggressive parameter filtering is less important when pseudo-label quality is already high.

\subsection{Hyperparameter Sensitivity}

\begin{figure}[H]
    \centering
    \includegraphics[hiresbb,width=\columnwidth]{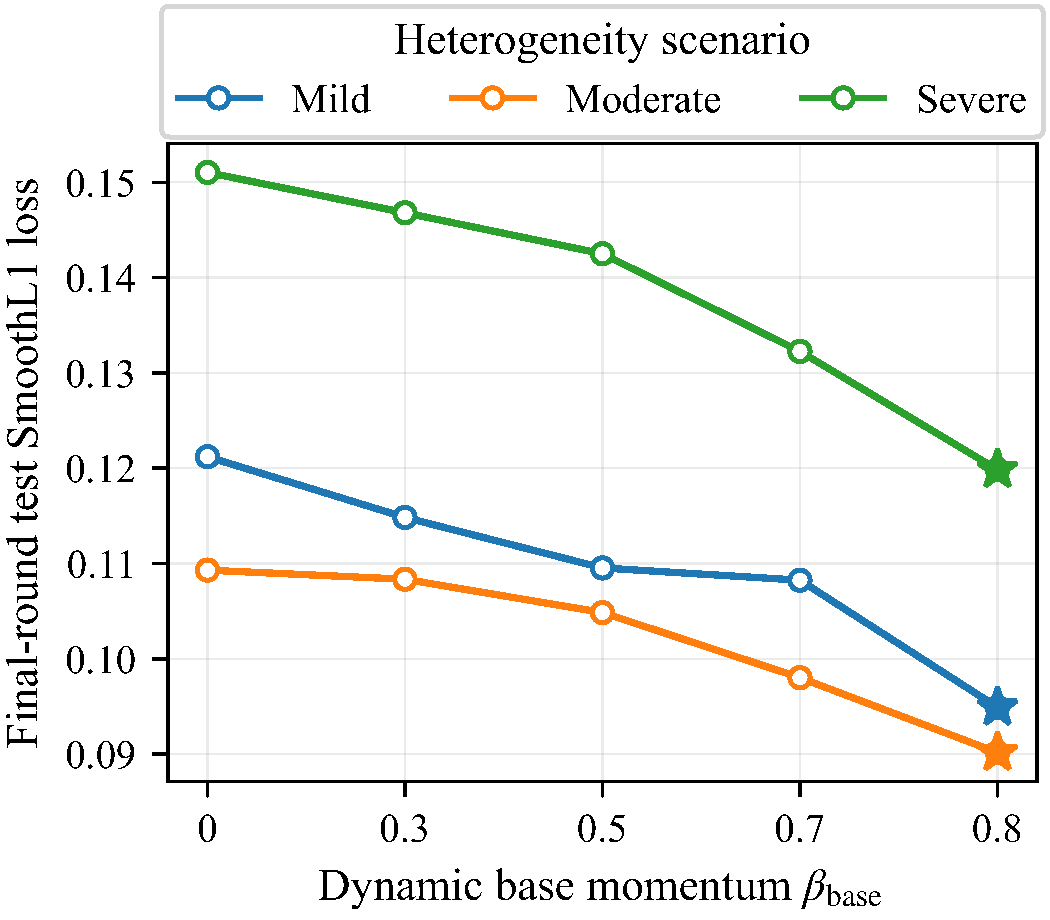}
    \caption{Sensitivity to the T-SMA base momentum with fixed $\rho=0.1$ and $\Delta\beta=0.18$. Stars mark the best tested value in each scenario.}
    \label{fig:beta_sensitivity}
\end{figure}

Fig.~\ref{fig:beta_sensitivity} evaluates the effect of the base momentum $\beta_{\mathrm{base}}$ on test Smooth L1 loss with $\rho=0.1$ and $\Delta\beta=0.18$. Increasing $\beta_{\mathrm{base}}$ from 0 to 0.8 lowers the final loss in all three scenarios. The tested upper bound is 0.8 because the final momentum $\beta_{\mathrm{base}}+\Delta\beta$ must remain below 1. With $\Delta\beta=0.18$, $\beta_{\mathrm{base}}=0.8$ yields a final momentum of 0.98. At $\beta_{\mathrm{base}}=0.8$, the mild, moderate, and severe losses are 0.0949, 0.0901, and 0.1198. This trend confirms that stronger late-stage inertia is beneficial when client optima are separated by data heterogeneity.

\subsection{Convergence Time}

\begin{figure}[!t]
    \centering
    \includegraphics[hiresbb,width=\columnwidth]{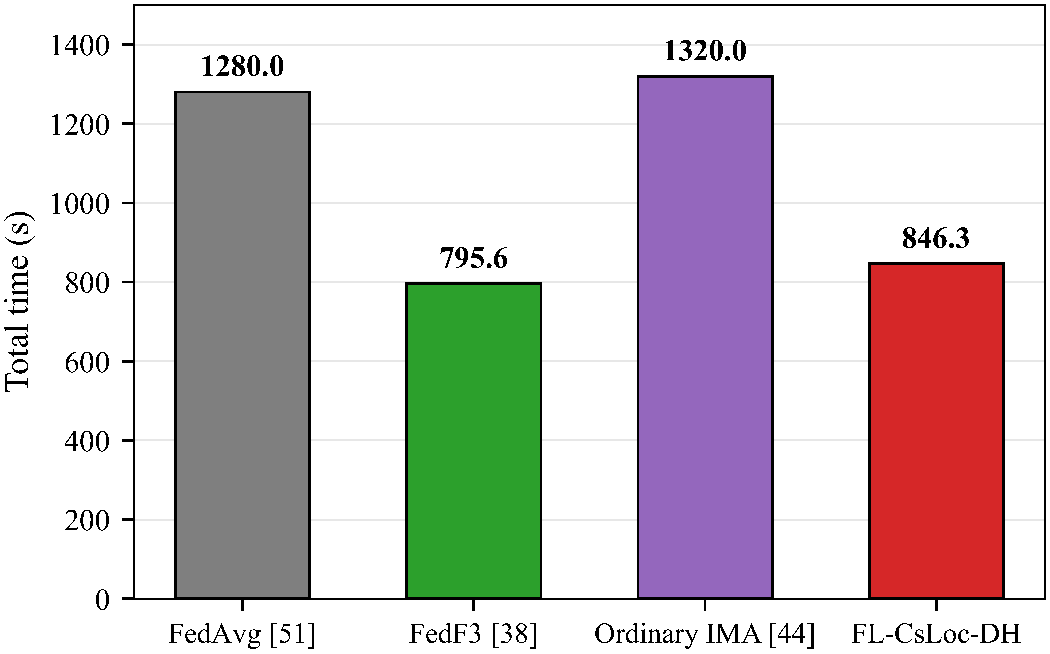}
    \caption{Estimated total time to convergence, including communication delay and local computation.}
    \label{fig:convergence_time}
\end{figure}

Fig.~\ref{fig:convergence_time} compares estimated end-to-end time to convergence. FL-CsLoc-DH requires 846.3 s, compared with 1280.0 s for FedAvg and 1320.0 s for ordinary IMA \cite{zhou2024understanding}, corresponding to reductions of 33.9\% and 35.9\%. FedF3 \cite{peng2025navigating} is faster at 795.6 s because hard parameter selection reduces its per-round workload. FL-CsLoc-DH incurs 6.4\% more time than FedF3, but provides lower final loss under the main dual heterogeneity setting. The result shows that the proposed soft reliability mechanisms trade a small amount of runtime for improved accuracy and stability while remaining substantially faster than FedAvg and ordinary IMA.

\section{Conclusion}
This paper presented FL-CsLoc-DH for FL-based crowdsensing localization under the inevitable data and annotation heterogeneity. The proposed framework follows the localization model update path from client training to server aggregation: TS-NGR reduces the influence of degraded client updates, S-FCA softly retains parameter increments according to labeled and unlabeled Fisher evidence, and T-SMA stabilizes the late-stage global localization model trajectory. Experiments on the ultra dense indoor MaMIMO CSI dataset showed that these mechanisms provide complementary accuracy and convergence gains under isolated and dual heterogeneity. The results indicate that jointly controlling client reliability, parameter reliability, and trajectory stability is effective for privacy-preserving localization when clients differ in propagation environment and label availability.

\ifCLASSOPTIONcaptionsoff
  \newpage
\fi

\bibliographystyle{IEEEtran}
\bibliography{IEEEfull,cite}
\end{document}